\documentclass{elsart}
\usepackage{graphicx}
\usepackage{amssymb}
\usepackage{epsfig}
\usepackage{epsf}
\usepackage{amsmath}
\usepackage{subfigure}

\begin{document}

\begin{frontmatter}

\title{Genus Distributions For Extended Matrix Models Of RNA}

\author{I. Garg,}
\author{N. Deo\corauthref{cor}}
\corauth[cor]{Corresponding author.}
 \ead{ndeo@physics.du.ac.in}
\address{Department of Physics and Astrophysics,\\
University of Delhi, Delhi 110007, India.\\}

\begin{abstract}
We construct and study an extended random matrix model of RNA
(polymer) folding. A perturbation which acts on all the nucleotides
in the chain is added to the action of the RNA partition function.
The effect of this perturbation on the partition function and the
Genus Distributions is studied. This perturbation distinguishes
between the paired and unpaired bases. For example, for $\alpha = 1$
(where $\alpha$ is the ratio of the strengths of the original and
perturbed term in the action) the partition function and genus
distribution for odd lengths vanish completely. This partition
function and the genus distribution is non-zero for even lengths
where structures with fully paired bases only remain. This implies
that (i). the genus distributions are different and (ii). there is a
``structural transition'' (from an ``unpaired-paired base phase'' to
a ``completely paired base phase'') as $\alpha$ approaches 1 in the
extended matrix models. We compare the results of the extended RNA
model with the results of G. Vernizzi, H. Orland and A. Zee in PRL
94, 168103(2005).
\end{abstract}

\end{frontmatter}

\section{Introduction}
\label{1.}

RNA (Ribose Nucleic Acid) is the only known biomolecule which plays
the dual role of being a carrier of genetic information and an
enzyme in important biological reactions \cite{2}. RNA carries the
genetic information from DNA to proteins (where the process of
transfer of information from DNA to RNA is referred to as
Transcription and from RNA to proteins is called Translation). Over
the years, the discovery of enzymatic and other functional roles of
ribozymes (larger RNA's $\sim$ 3000 bases long) have strengthened
the importance of RNA in cellular functions. So a study of the RNA
structure, classification of different levels of structure,
complexity at each level, the energetics and stability have captured
the interest of scientists from all genres.

Structurally, bio-polymers have been classified into three
hierarchical levels which are graded in increasing levels of
complexity and instability (in the given order):\\
(i). Primary Structure : the chemical sequence or the sequence of
nucleotides (planar diagrams), (ii). Secondary Structure : the local
short-range pairing of the complementary Watson-Crick bases A-U and
G-C via hydrogen bonding (planar diagrams) and (iii). Tertiary
structure : the spatial arrangement of secondary units by means of
several van der Waals contacts, hydrogen bonding between the
complementary base pairs and the interactions in which loops and
bulges can themselves partially pair leading to the formation of
pseudoknots (non planar diagrams). Pseudoknots are conformations
whose associated disk diagrams aren't planar \cite{3} or we can say
that pseudoknots are false knots which have higher genii (a genus
may be understood as the number of handles that a surface has so
that the arcs, representing pairings between the bases, do not
intersect) and are much more complex than the previous two levels of
structures. Most of the work in the field of RNA folding has been
carried out on secondary structures. Monte Carlo techniques and
numerous algorithms are available to solve problems related to
secondary structures \cite{3,4,5,6}. Very little, though is known
about the tertiary structure of RNA because of the complexity
associated with the structures. Immense energies are required to
fold RNA into their folded conformations. The energies involved are
of the order of $\sim$100 Kcal/mol or more \cite{7,8}. In RNA's, the
complexity increases by manifolds when one goes from secondary to
tertiary structures. Folding of the secondary structures to from
tertiary structures is largely governed by the presence of $Mg^{2+}$
ions as demonstrated by experimental studies. The biological
function of RNA is governed by these folded structures i.e the
tertiary structures. So in order to understand the role that RNA's
play in biological processes it becomes essential to understand
their folded conformations.

Matrix models have proved to be useful and important in areas such
as disordered condensed matter systems and quantum chromodynamics
among others. These models have been extremely useful in modeling
very complex systems. A random matrix model has been proposed in
\cite{1,9} to count and classify all possible structures of RNA
which can exist for a given length of the nucleotide chain.
However, the number of structures that have been discovered are a
very small subset of these vast number of structures. This may be
due to the constantly varying conditions inside a living cell. The
environment of a cell is subject to changing conditions of
temperature, salinity, acidity etc. which act as perturbations to
the processes taking place in a cell. These perturbations affect the
overall functioning of the cell. So the addition of a perturbation
in the potential of the original model of RNA \cite{1,9} and to
observe its effect is very important. In this work a very simple
perturbation is studied which does not drastically change the
essence of the model proposed in \cite{1,9}. However, it brings
changes in the enumeration of structures of RNA. In particular, for
a certain strength of this perturbation a phase of RNA structures
with no unpaired bases is found. This will result in a limited
activity of the RNA due to the paired conformations alone. Thus a
seemingly trivial change in the mathematical structure of the RNA
model produces structures of RNA which leads to a non-trivial change
in their biological activity.


The random matrix model of RNA folding proposed in \cite{9} is a
theoretical model which focuses on the topological aspect of the RNA
folding and is constructed as an (N$\times$N) matrix field theory.
In the model, a matrix $V_{i, j}$ is considered which accounts for
the attractive energies between the nucleotides, disregards any self
interaction of the nucleotides, maintains finite flexibility of the
nucleotide chain and the fact that only those nucleotides which are
at least 4 nucleotides apart can interact \cite{9}. The model makes
use of the notations of quantum chromodynamics in terms of the
Feynmann diagrams. In the large N limit, matrix models can be
expanded in powers of $\frac{1}{N^{2}}$ which also gives a
topological expansion as was originally observed by G 't Hooft
\cite{10}. This allows us to identify the tertiary structures as
terms with powers of $\frac{1}{N^{2}}$. The RNA matrix model in
\cite{1} enumerates all the secondary structures with pseudoknots by
making the following simplifications: the polymer chain is assumed
to be infinitely flexible so that nucleotides less than 4
nucleotides apart can also interact, sterical constraints are
neglected so that any base pairing is allowed irrespective of the
type of nucleotide and the interaction between the bases in the
chain is assumed to be simple with all the base-pairings taking
place with a similar strength `v'. This means that the interaction
matrix $V_{i, j}$ (which is a (L$\times$L) matrix, L being the
length of the polymer chain) has all the elements equal to `v'.

In this paper, section 2 introduces the extended matrix model where
a perturbation is added to all the bases in the nucleotide chain
keeping the same simplifications as in \cite{1}. A parameter
$\alpha$ is found which can take different values and hence it
distinguishes between the different generalized models. A general
form of the partition function for the extended matrix model is
found for any length L and $\alpha$. We also discuss the extended
matrix model with $\alpha=1$. In section 3 we discuss the
diagrammatic representation of the extended matrix model
corresponding to any $\alpha$. Section 4 deals with the genus
distributions (${\cal N}_{\alpha}$ and $a_{L, g, \alpha}$) of the
extended matrix model, where the odd lengths vanish in the partition
function $Z_{L, \alpha}(N)$, as $\alpha$ approaches 1. We will refer
to the RNA folding model in \cite{1,9} as the RNA-MM and the
extended matrix model as RNA-EMM from now on in the literature.
\section{Extended Matrix Model of RNA (RNA-EMM) with a Perturbation on all the Bases}
\label{2.}

We propose an extension of the RNA-MM discussed in \cite{1}. Here we
add a perturbation term in the action of the partition function of
the nucleotide chain. The partition function of the extended matrix
model is,
\begin{eqnarray}
 Z_{L, \alpha}(N) & = & \frac{1}{A_{L}(N)} \int \prod_{i=1}^L d\phi_{i}
                        exp^{-\frac{N}{2} \sum_{i, j = 1}^L (V_{i, j})^{-1} Tr \phi_{i}
                        \phi_{j}} exp^{(-N) \sum_{i = 1}^L (W_{i})^{-1} Tr \phi_{i}}\nonumber\\
                  &   & \frac{1}{N} Tr \prod_{i = 1}^L (1 + \phi_{i})
\end{eqnarray}

where $\phi_{i}$ are L independent (N$\times$N) hermitian matrices,
$V_{i, j}$ is an (L$\times$L) interaction matrix containing
interactions between the nucleotides, L being the length of the
nucleotide chain, $\prod_{i} (1+\phi_{i})$ is an ordered product
over $\phi_{i}$'s and $A_{L}(N)$ is the normalization constant given
by,
\begin{equation}
A_{L}(N) = \int \prod_{i = 1}^{L} d\phi_{i} exp^{-\frac{N}{2}
\sum_{i, j = 1}^L (V_{i, j})^{-1} Tr \phi_{i} \phi_{j}} exp^{(-N)
\sum_{i = 1}^L (W_{i})^{-1} Tr \phi_{i}}.
\end{equation}

These RNA matrix models (both RNA-MM and RNA-EMM) are variants of
Gaussian Penner Models. Addition of a linear term in this action is
non-trivial as shown in references \cite{11,12} where a linear term
$\phi$ is added to the potential $V(\phi) = g\phi^{2} + \mu\phi^{4}$
of the partition function for the multi-cut matrix model.

The perturbation term in the partition function of equation (1)
is,
\begin{equation}
exp^{(-N) \sum_{i = 1}^L (W_{i})^{-1} Tr \phi_{i}}
\end{equation}

where $W_{i} = w$, $w$ being the strength of the perturbation. We
consider $V_{i, j} = v$. The normalization constant $A_{L}(N)$ can
be written as $A_{L}(N)$ = $exp^{\frac{N}{2} Tr (\frac{v}{w^{2}})} \\
\int \prod_{i=1}^{L} d\Phi_{i} exp^{-\frac{N}{2} Tr (\Phi_{i} (V_{i,
j})^{-1} \Phi_{j})}$ where $\Phi_{i} = (\phi_{i} + V_{i,
j}W^{-1}_{j})$. Carrying out a series of Hubbard Stratonovich
Transformations, the integral in equation (1) is reduced to an
integral over a single (N$\times$N) matrix $\sigma$,

\begin{equation}
Z_{L, \alpha}(N) = \frac{1}{R_{L}(N)} \int d\sigma
exp^{-\frac{N}{2v} Tr (\sigma + \frac{v}{w})^{2}} \frac{1}{N} Tr
(1 + \sigma)^L
\end{equation}

where $R_{L}(N) = \int d\sigma exp^{-\frac{N}{2v} Tr (\sigma +
\frac{v}{w})^{2}}$. We make a redefinition such that $\sigma^\prime
= (\sigma + \frac{v}{w}) = (\sigma + \alpha)$ where we define
$\alpha = \frac{v}{w}$, the ratio of the strength of interaction
between the nucleotides to the strength of the perturbation. We
therefore get from equation (4),

\begin{equation}
Z_{L, \alpha}(N) = \frac{1}{R_{L}(N)} \int d\sigma^\prime
exp^{-\frac{N}{2v} Tr (\sigma^\prime)^2} \frac{1}{N} Tr (1 +
\sigma^\prime - \alpha)^L.
\end{equation}

We now introduce the spectral density $\rho_{N, \alpha}(\lambda)$ of
a Gaussian matrix model defined at finite N as,
\begin{equation}
\rho_{N, \alpha}(\lambda) = \frac{1}{R_{L}(N)} \int d\sigma^\prime
exp^{-\frac{N}{2v} Tr (\sigma^\prime)^2} \frac{1}{N} Tr
\delta(\lambda - \sigma^\prime).
\end{equation}

Making use of the identity $\int_{-\infty}^{+\infty} d\lambda
\rho_{N, \alpha}(\lambda) = 1$ in equation (5) the partition
function can be written as,

\begin{equation}
Z_{L, \alpha}(N) = \int_{-\infty}^{+\infty} d\lambda \rho_{N,
\alpha}(\lambda) (1 + \lambda - \alpha)^L.
\end{equation}

Defining $G(t, N, \alpha)$ as the exponential generating function
of $Z_{L, \alpha}(N)$ we can write,
\begin{equation}
G(t, N, \alpha) \equiv \sum_{L = 0}^{\infty} Z_{L, \alpha}(N)
\frac{t^L}{L!} = \int_{-\infty}^{+\infty} d\lambda \rho_{N,
\alpha}(\lambda) \sum_{L = 0}^{\infty} \frac{t^L (1 + \lambda -
\alpha)^L}{L!}.
\end{equation}

\begin{equation}
G(t, N, \alpha) = \int_{-\infty}^{+\infty} d\lambda \rho_{N,
\alpha}(\lambda) exp^{t (1 + \lambda - \alpha)}.
\end{equation}

We use the form of spectral density $\rho_{N, \alpha}(\lambda)$ from
\cite{13,14},
\begin{equation}
\rho_{N, \alpha}(\lambda) = \frac{exp^{-\frac{N
(\lambda)^2}{2v}}}{\sqrt{2 \Pi v N}} \sum_{k = 0}^{N - 1}
\binom{N}{k+1} \frac{H_{2k}(\lambda \sqrt{\frac{N}{2v}})}{2^k k!}
\end{equation}

where $H_{2k}(\lambda \sqrt{\frac{N}{2v}})$ represents Hermite
polynomials. Using equation (10) the generating function $G(t, N,
\alpha)$ in equation (9) can be written as,
\begin{eqnarray}
G(t, N, \alpha) & = & \frac{1}{\sqrt{2 \Pi v N}} \sum_{k = 0}^{N -
                      1} \binom{N}{k+1} \frac{1}{2^k k!}\int_{-\infty}^{+\infty} d\lambda
                      exp^{-\frac{N(\lambda)^{2}}{2v}} exp^{t(\lambda + 1 -
                      \alpha)}\nonumber\\
                &   & H_{2k}(\lambda \sqrt{\frac{N}{2v}}).
\end{eqnarray}

Rewriting the above equation,
\begin{eqnarray}
G(t, N, \alpha) & = & \frac{exp^{\left[\frac{v t^2}{2 N} + t(1 -
                      \alpha)\right]}}{\sqrt{2 \Pi v N}} \sum_{k = 0}^{N -
                      1} \binom{N}{k+1} \frac{1}{2^k k!} \int_{-\infty}^{+\infty} d\lambda exp^{-\left[\sqrt{\frac{N}{2 v}}\lambda - \sqrt{\frac{2
                      v}{N}}\frac{t}{2}\right]^{2}} \nonumber\\
                &   & H_{2k}(\lambda \sqrt{\frac{N}{2v}}).
\end{eqnarray}

Using a standard result of integration over Hermite polynomials
\cite{15},
\begin{equation}
\int_{-\infty}^{+\infty} dx\hspace{0.2cm}exp^{-(x-y)^2} H_{n}(x) =
\sqrt{\Pi} y^n 2^n
\end{equation}

equation (12) solves to,
\begin{equation}
G(t, N, \alpha) = exp^{\frac{v t^2}{2N} + t(1 - \alpha)}
\left[\frac{1}{N} \sum_{k = 0}^{N - 1} \binom{N}{k+1} \frac{(v
t^2)^k}{k! N^k}\right].
\end{equation}

\begin{table*}
\caption{The Table lists the partition functions $Z_{L,
\alpha}(N)$ for $\alpha = (0.25,0.5)$ for different lengths L.}
\vspace{0.5cm}
\begin{tabular}{|l|l|l|}
\hline
L & $Z_{L, \alpha}(N)\left[\alpha = 0.25\right]$ & $Z_{L, \alpha}(N)\left[\alpha = 0.5\right]$ \\[0.5ex]
\hline
1 & $3/4$ & $1/2$\\
\hline
2 & v+$9/16$ & v+$1/4$\\
\hline
3 & $9v/4$+$27/64$ & $3v/2$ + $1/8$\\
\hline
4 & $81/256$+$27v/8$+2$v^2$+$v^2/N^2$ & $1/16$+$3v/2$+2$v^2$+$v^2/N^2$\\
\hline
5 & $243/1024$+$135v/32$+$15v^2/2$+ & $1/32$+$5v/4$+5$v^2$+$5v^2/2N^2$\\
  & $15v^2/4 N^2$                   & \\
\hline
6 & $729/4096$+$1215v/256$+$135v^2/8$+ & $1/64$+$15v/16$+$15v^2/2$+$15v^2/4N^2$+5$v^3$+$10v^3/N^2$ \\
  & $135v^2/16 N^2$+5$v^3$+10$v^3/N^2$ & \\
\hline
7 & $2187/16384$+$5103v/1024$+$945v^2/32$+ & $1/128$+$21v/32$+$35v^2/4$+$35v^2/8N^2$+$35v^3/2$+$35v^3/N^2$\\
  & $945v^2/64 N^2$+$105v^3/4$+$105v^3/2 N^2$ & \\
\hline
8 & $6561/65536$+$5103v/1024$+$2835v^2/64$+ & $1/256$+$7v/16$+$35v^2/4$+$35v^2/8N^2$+35$v^3$+\\
  & $2835v^2/128N^2$+$315v^3/4$+$315v^3/2N^2$ & $70v^3/N^2$+14$v^4$+$21v^4/N^4$+$70v^4/N^2$\\
  & +14$v^4$+21$v^4/N^4$+70$v^4/N^2$          & \\
\hline
\end{tabular}
\end{table*}

%

Combining equations (8) and (14) we have,
\begin{equation}
G(t, N, \alpha) \equiv \sum_{L = 0}^{\infty} Z_{L, \alpha}(N)
\frac{t^L}{L!} = exp^{\frac{v t^2}{2N} + t(1 - \alpha)} \left[
\frac{1}{N} \sum_{k = 0}^{N - 1} \binom{N}{k+1} \frac{(t^2
v)^k}{k! N^k} \right].
\end{equation}

We observe from equation (15) that by taking different values of
$\alpha$ (i.e. by varying the ratio of the strengths of interaction
between the bases and the perturbation in the action of the
partition function) we can have several extensions of the same
model. For instance, for $\alpha = 0$, we have the RNA-MM \cite{1}.
We substitute $\alpha = 0.25, 0.5, 0.75$ in equation (15) and find
their corresponding partition functions ($\alpha=0.25, 0.5$ are
listed in Table 1). For completeness we write the explicit
dependence of the partition function $Z_{L, \alpha}(N)$ on N in
terms of (the topological parameter), genus g, as in \cite{1},
$Z_{L, \alpha}(N) = \sum_{g = 0}^{\infty} a_{L, g, \alpha}
\frac{1}{N^{2g}}$. Here, the coefficients $a_{L, g, \alpha}$'s give
the number of diagrams (structures) at a given length L, genus g and
$\alpha$. We also define the total number of diagrams as ${\cal
N}_{\alpha}$, for a particular length and $\alpha$, irrespective of
the genus.

\begin{table*}
\caption{The Table lists the partition functions $Z_{L, \alpha}(N)$
for any $\alpha$ and for different lengths L.}
\vspace{0.5cm}
\begin{tabular}{|l|l|}
\hline
L & $Z_{L, \alpha}(N)\left[RNA-EMM\right]$\\[0.5ex]
\hline
1 & 1-$\alpha$\\
\hline
2 & 1+v+$(\alpha)^{2}$-2$(\alpha)$\\
\hline
3 & 1+3v-$(\alpha)^{3}$+3$(\alpha)^{2}$-3$(\alpha)$-3v$(\alpha)$\\
\hline
4 & 1+6v+2$v^2$+$v^2/N^2$+$(\alpha)^{4}$-4$(\alpha)^{3}$+6$(\alpha)^{2}$+6v$(\alpha)^{2}$-4$(\alpha)$-12v$(\alpha)$\\
\hline
5 & 1-$(\alpha)^{5}$+5$(\alpha)^{4}$-10$(\alpha)^{3}$+10$(\alpha)^{2}$-5$(\alpha)$+10v+10$v^2$+5$v^2/N^2$-10v$(\alpha)^{3}$+30v$(\alpha)^{2}$-\\
  & 30v$(\alpha)$-10$v^2(\alpha)$-5$v^2(\alpha)/N^2$\\
\hline
6 & 1+$(\alpha)^{6}$-6$(\alpha)^{5}$+15$(\alpha)^{4}$-20$(\alpha)^{3}$+15$(\alpha)^{2}$-6$(\alpha)$+15v+30$v^2$+15$v^2/N^2$+5$v^3$+10$v^3/N^2$+15v$(\alpha)^{4}$-\\
  & 60v$(\alpha)^{3}$+90v$(\alpha)^{2}$+30$v^2(\alpha)^{2}$+15$v^2(\alpha)^{2}/N^2$-60v$(\alpha)$-60$v^2(\alpha)$-30$v^2(\alpha)/N^2$\\
\hline
7 &  1-$(\alpha)^{7}$+7$(\alpha)^{6}$-21$(\alpha)^{5}$+35$(\alpha)^{4}$-35$(\alpha)^{3}$+21$(\alpha)^{2}$-7$(\alpha)$+21v+70$v^2$+\\
  &  35$v^2/N^2$+35$v^3$+70$v^3/N^2$-21v$(\alpha)^{5}$+105v$(\alpha)^{4}$-210v$(\alpha)^{3}$-70$v^2(\alpha)^{3}$-35$v^2(\alpha)^{3}/N^2$+210v$(\alpha)^{2}$+\\
  &  210$v^2(\alpha)^{2}$+105$v^2(\alpha)^{2}/N^2$-105v$(\alpha)$-210$v^2(\alpha)$-105$v^2(\alpha)/N^2$-35$v^3(\alpha)$-70$v^3(\alpha)/N^2$\\
\hline
\end{tabular}
\end{table*}

\textit{\underline{The General Form of $Z_{L, \alpha}(N)$}}: The
general form of the partition function for the extended matrix model
can be obtained from equation (1) after completing the square in the
exponent of the numerator and in the normalization constant
$A_{L}(N)$ in terms of the variable $\Phi$ as,

\begin{equation}
Z_{L, \alpha}(N) = \frac{1}{\tilde {A}_{L}(N)} \int
\prod_{i=1}^{L} d\Phi_{i} exp^{-\frac{N}{2} Tr (\Phi_{i} (V_{i,
j})^{-1} \Phi_{j})} \frac{1}{N} Tr \prod_{i = 1}^{L} (1 +
\Phi_{i} - \alpha).
\end{equation}

The general form of $Z_{L, \alpha}(N)$ for the extended matrix model
from equation (16) via Wick Theorem is,

\begin{eqnarray}
Z_{L, \alpha}(N) & = & (1 - \alpha)^{L} + (1 - \alpha)^{(L - 2)}
                       \sum_{i < j} V_{i, j} + (1 - \alpha)^{(L - 4)} \sum_{i < j < k
                       < l} V_{i, j} V_{k, l} + \nonumber\\
                 &   & (1 - \alpha)^{(L - 4)} \sum_{i < j <k < l} V_{i, l} V_{j, k}
                       + (\frac{1}{N})^2 (1 - \alpha)^{(L - 4)} \sum_{i < j < k < l} V_{i, k} V_{j, l}\nonumber\\
                 &   & +............
\end{eqnarray}

We observe that the general form of $Z_{L, \alpha}(N)$ has the same
structure as in \cite{1} with an additional multiplicative factor of
powers of $(1 - \alpha)$ on each term (Table 2 lists $Z_{L,
\alpha}(N)$ upto $L = 7$ in terms of $\alpha$'s). The partition
functions can be obtained from equation (17) by substituting the
desired $\alpha$ for a particular length.

\begin{table*}
\caption{The Table lists the partition functions $Z_{L, \alpha}(N)$
for different lengths L of the nucleotide chain for $\alpha=0$
(RNA-MM) and $\alpha=1$ (RNA-EMM).} \vspace{0.5cm}
\begin{tabular}{|l|l|l|}
\hline L & $Z_{L, \alpha=0}(N)\left[RNA-MM\right]$ &
$Z_{L, \alpha=1}(N)\left[RNA-EMM\right]$ \\[0.5ex]
\hline
1 & 1 & 0 \\
\hline
2 & 1+v & v \\
\hline
3 & 1+3v & 0 \\
\hline
4 & 1+6v+2$v^2$+$v^2/N^2$ & 2$v^2$+$v^2/N^2$ \\
\hline
5 & 1+10v+10$v^2$+$5v^2/N^2$ & 0 \\
\hline
6 & 1+15v+30$v^2$+5$v^3$+$(15v^2+10v^3)/N^2$ & 5$v^3$+$10v^3/N^2$ \\
\hline
7 & 1+21v+70$v^2$+35$v^3$+$(35v^2+70v^3)/N^2$ & 0 \\
\hline
8 & 1+28v+140$v^2$+140$v^3$+14$v^4$+21$v^4/N^4$+ & 14$v^4$+70$v^4/N^2$+21$v^4/N^4$\\
  & (70$v^2$+280$v^3$+70$v^4$)/$N^2$             & \\
\hline
\end{tabular}
\end{table*}

\textit{\underline{Extended Model With $\alpha=1$}}: We now
substitute $\alpha = 1$ in equation (15) i.e. the perturbation is
acting on all the bases in the nucleotide chain with a strength the
same as the interaction between the bases. We get,

\begin{equation}
G(t, N, \alpha) \equiv \sum_{L = 0}^{\infty} Z_{L, \alpha}(N)
\frac{t^L}{L!} = exp^{\frac{v t^2}{2N}} \left[ \frac{1}{N}
\sum_{k = 0}^{N - 1} \binom{N}{k+1} \frac{(t^2 v)^k}{k! N^k}
\right].
\end{equation}

We can see from equation (18) that only even length partition
functions will be non-zero whereas for odd lengths, the partition
function will vanish (results are summarized in Table 3). It will be
shown in the next section that $Z_{L, \alpha=1}(N)=0$ for odd
lengths implies that no unpaired base is allowed for the $\alpha=1$
phase. This suggests a considerable change in the genus
distributions for the extended matrix model especially for the case
when $\alpha=1$. This is investigated in detail in section 4.

\section{Diagrammatic Representation of the Extended Model}
\label{3.}

We explain in this section the physical interpretation of the
general form of $Z_{L, \alpha}(N)$ obtained in section 2 for the
extended matrix model. The general form of the partition function
for the extended model is given by equation (17). Each term in the
partition function is accompanied by a factor of powers of $(1 -
\alpha)$. The partition function corresponding to $L=4$ (Table 2) is
given by $Z_{L=4, \alpha}(N) =
1+6v+2v^{2}+v^{2}/N^{2}+\alpha^{4}-4\alpha^{3}+6\alpha^{2}+6v\alpha^{2}-4\alpha-12v\alpha$
or simply $Z_{L=4, \alpha}(N) =
(1-\alpha)^{4}+6v(1-\alpha)^{2}+2v^{2}+v^{2}/N^{2}$. This represents
a total of 10 diagrams (Figure 1). In the diagrammatic
representation in these models (using quantum chromodynamics), the
RNA backbone is represented by a solid line (representing the quark
propagator) and the hydrogen bonds between different nucleotides as
dotted arcs (representing the gluon propagators) joining two
nucleotides on the solid line. Therefore, a primary structure is
shown with the nucleotides (or the dots) on the solid line,
secondary structure is represented by the arcs (dotted lines) drawn
between nucleotides (dots) on the solid line (provided arcs do not
intersect) and tertiary structures are those secondary structures in
which the arcs intersect \cite{9}. In these diagrams, the powers of
`v' give the number of arcs in the corresponding Feynmann diagrams
(diagrams are given by the coefficients of the v term). These terms
correspond to the planar diagrams. The terms with powers of
$1/N^{2}$ represent tertiary structures with non-zero genus. The
first term in the partition function for $L=4$ is a planar term with
each unpaired base associated with the factor of $(1-\alpha)$, the
second term represents 6 diagrams with one arc each and each
unpaired base (2 in number) accompanied by $(1-\alpha)$, the third
term corresponds to two diagrams with two arcs each and no unpaired
bases at all and the last term represents one diagram with two
crossing arcs i.e. a tertiary term with genus one and no unpaired
bases. These diagrams are shown in Figure 1 with their respective
weights. The power of $(1-\alpha)$ gives the number of unpaired
bases in the diagram. However, for $\alpha=1$ the partition function
for $L=4$ is given by $Z_{L=4, \alpha=1}(N) = 2v^{2}+v^{2}/N^{2}$
i.e. a total of 3 diagrams (in Figure 1, the diagrams corresponding
to $v^{2}$ , $g=0$ and $g=1$). For this $\alpha$ value, only those
diagrams with completely paired bases remain. The diagrams with even
a single unpaired base are completely suppressed.

\begin{figure}
\includegraphics[width=8cm]{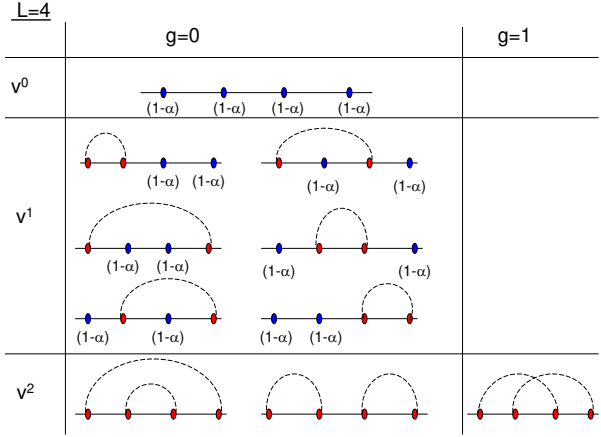}
\caption{The figure shows all the possible diagrams for $L=4$ for
the extended matrix model. The paired bases are shown to be
connected by arcs. Each unpaired base is associated with a factor of
$(1-\alpha)$.}
\end{figure}

We thus observe that for $\alpha < 1$, all kinds of structures with
fully unpaired, partially paired-partially unpaired and completely
paired bases exist. However, for $\alpha=1$, only structures with
completely paired bases exist. We refer to these structural
differences between the two regions $\alpha < 1$ and $\alpha=1$ as a
``structural transition'' from an ``unpaired-paired base phase'' to
a ``completely paired base phase'' respectively. Unpaired-paired
base phase comprises of all the structures enumerated by the RNA-MM
in \cite{1} with different weights associated with each unpaired
base for different $\alpha$'s. A completely paired base phase
consists of only those structures which have no unpaired bases at
all. For $L=4$ and any given $\alpha$, $a_{L=4, g=0} = 9$ and ${\cal
N}_{\alpha} = 10$, so $a_{L=4, g=0}/{\cal N}_{\alpha}=0.9$. This
value of the normalized distribution can be seen in Figure 4(a)
(i.e. corresponding to $g=0$). For $L=4$ and $\alpha=1$, $a_{L=4,
g=0} = 2$ and ${\cal N}_{\alpha} = 3$, so $a_{L=4, g=0}/{\cal
N}_{\alpha}=0.66$. This value of the normalized distribution can
also be seen in Figure 4(a).

We have seen that this perturbation has created a phase where RNA
structures with a limited biological activity (as only structures
with paired bases are possible) are separated out from the otherwise
possible vast number of structures.

\section{Genus Distributions for the Extended Matrix Models}
\label{4.}

We analyze the genus distributions for the extended matrix model and
compare them with the distributions of RNA-MM \cite{1}. We consider
in particular the extended matrix model with $\alpha=1$. For $\alpha
= 1$ the genus distributions are shown in Figure 2. Figure 2(a)
plots the normalized diagrams $a_{L, g, \alpha=1}/{\cal N}_{\alpha}$
Vs Length L at a fixed genus g. Figure 2(b) shows a plot of the
normalized diagrams $a_{L, g, \alpha=1}/{\cal N}_{\alpha}$ Vs g at a
particular length L. As indicated by the generating function in
equation (18), we find that the partition function for odd lengths
vanish (Table 3). The genus distributions are shown in Figure 2.

\begin{figure*}
\includegraphics*[width=7cm]{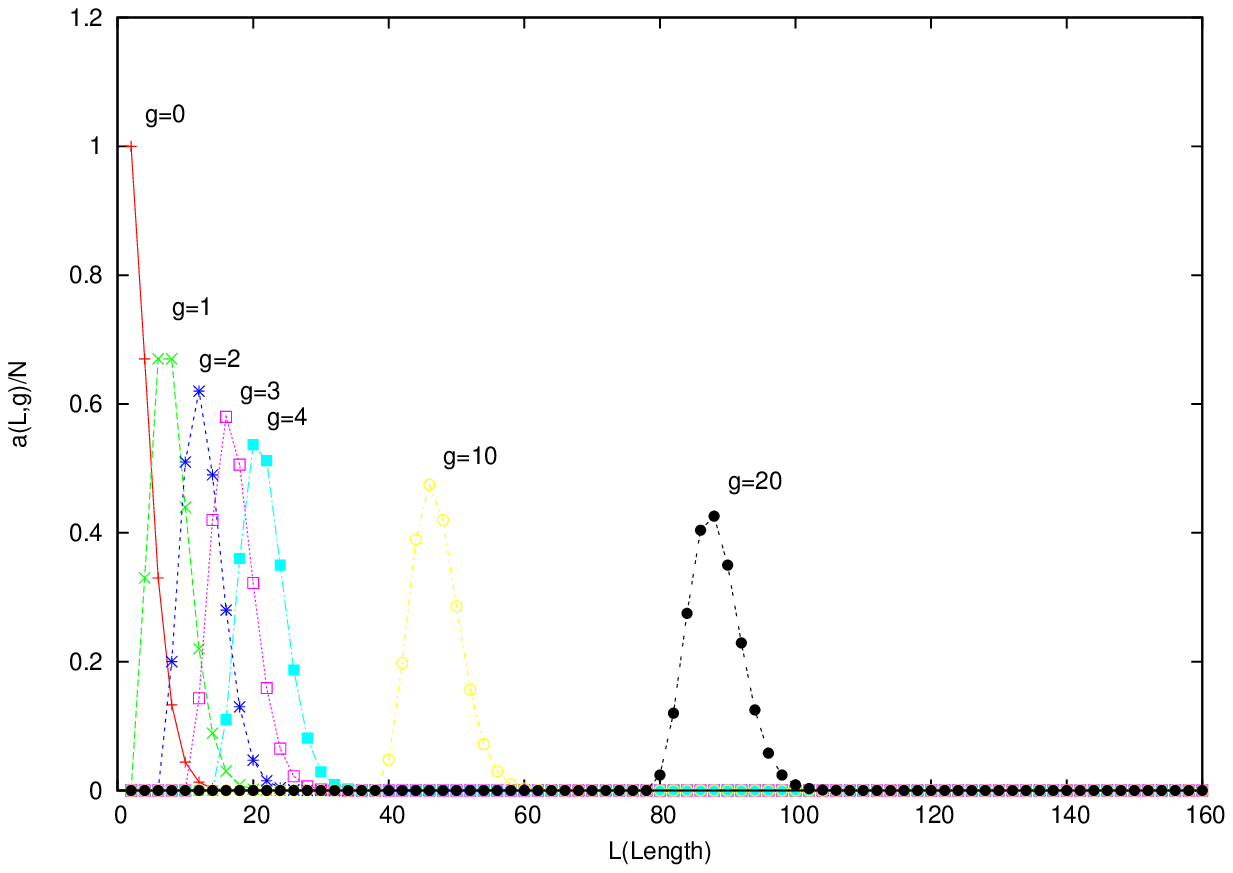}
\includegraphics*[width=7cm]{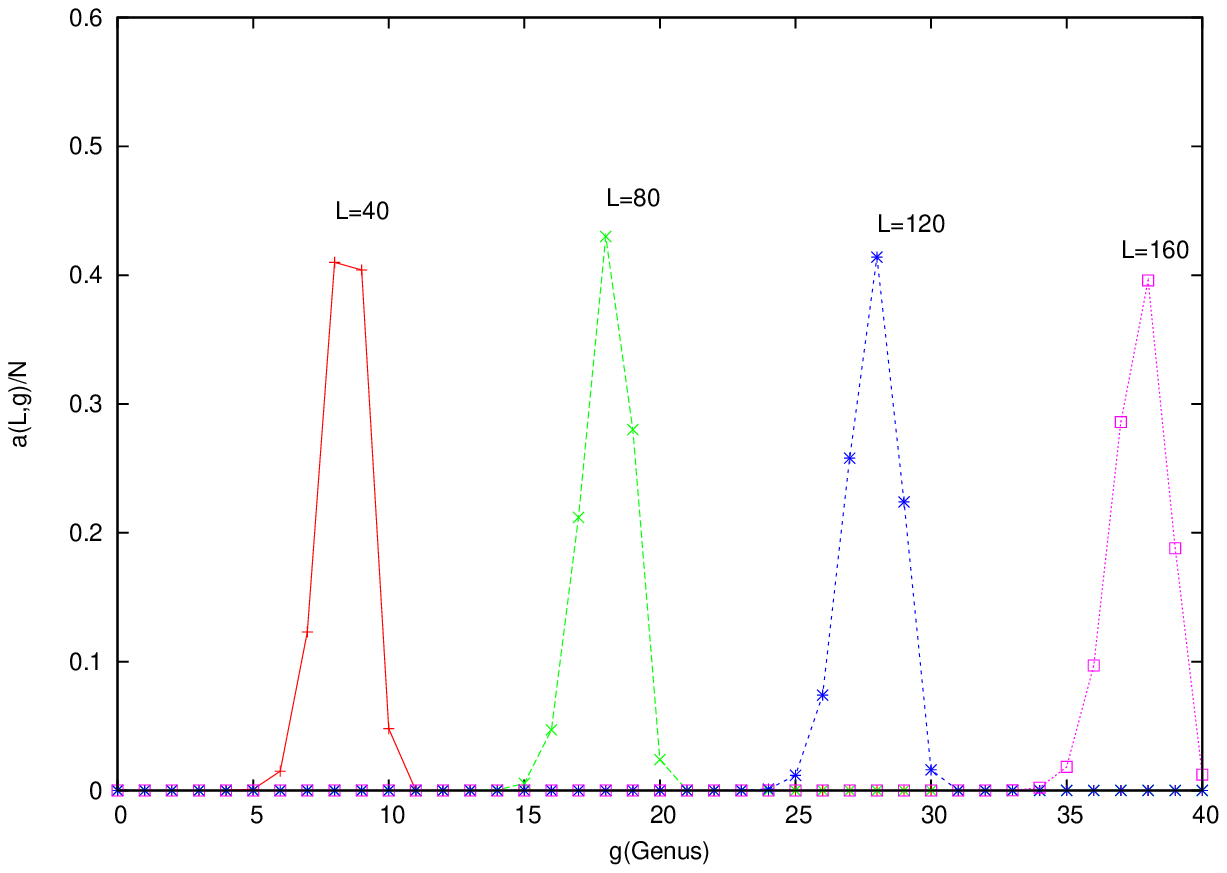}
\caption{(a) For $\alpha = 1$, the normalized diagrams, $a_{L, g,
\alpha}/{\cal N}_{\alpha}$ are plotted with respect to the
nucleotide chain lengths L, keeping the genus g fixed. Here, $a_{L,
g, \alpha}$ are the total number of diagrams at a given L, g and
$\alpha$ and ${\cal N}_{\alpha}$ are the total number of diagrams at
a particular L and $\alpha$, irrespective of g. (b) For $\alpha =
1$, the normalized diagrams $a_{L, g, \alpha}/{\cal N}_{\alpha}$ are
plotted as a function of genus. The length L of the chain is kept
fixed in order to obtain the genus distribution patterns for a
specific length. Lengths considered are L=40,80,120,160.
$\frac{a_{L, g}}{N}$ in the figures should be understood as
$\frac{a_{L, g, \alpha=1}}{{\cal N}_{\alpha}}$.}
\end{figure*}

\begin{figure*}
\includegraphics*[width=7cm]{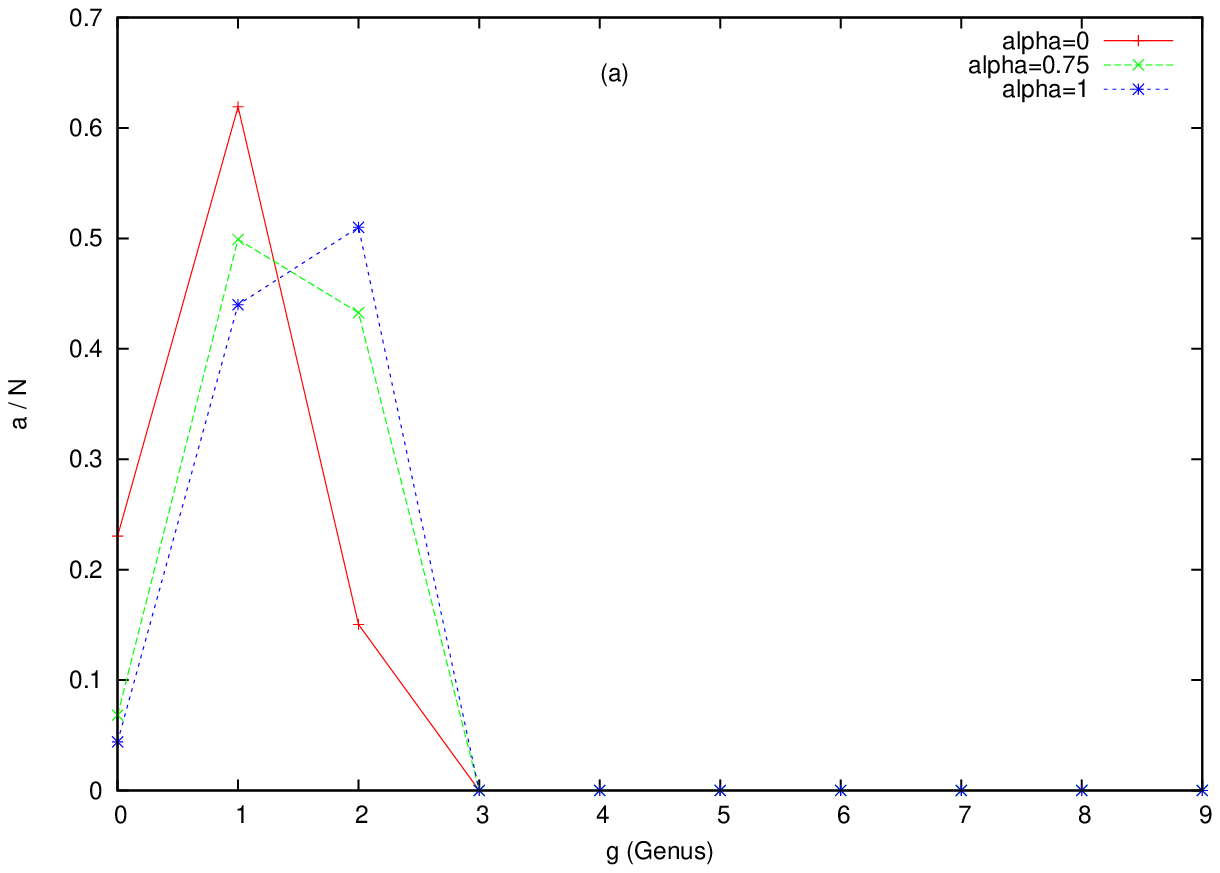}
\includegraphics*[width=7cm]{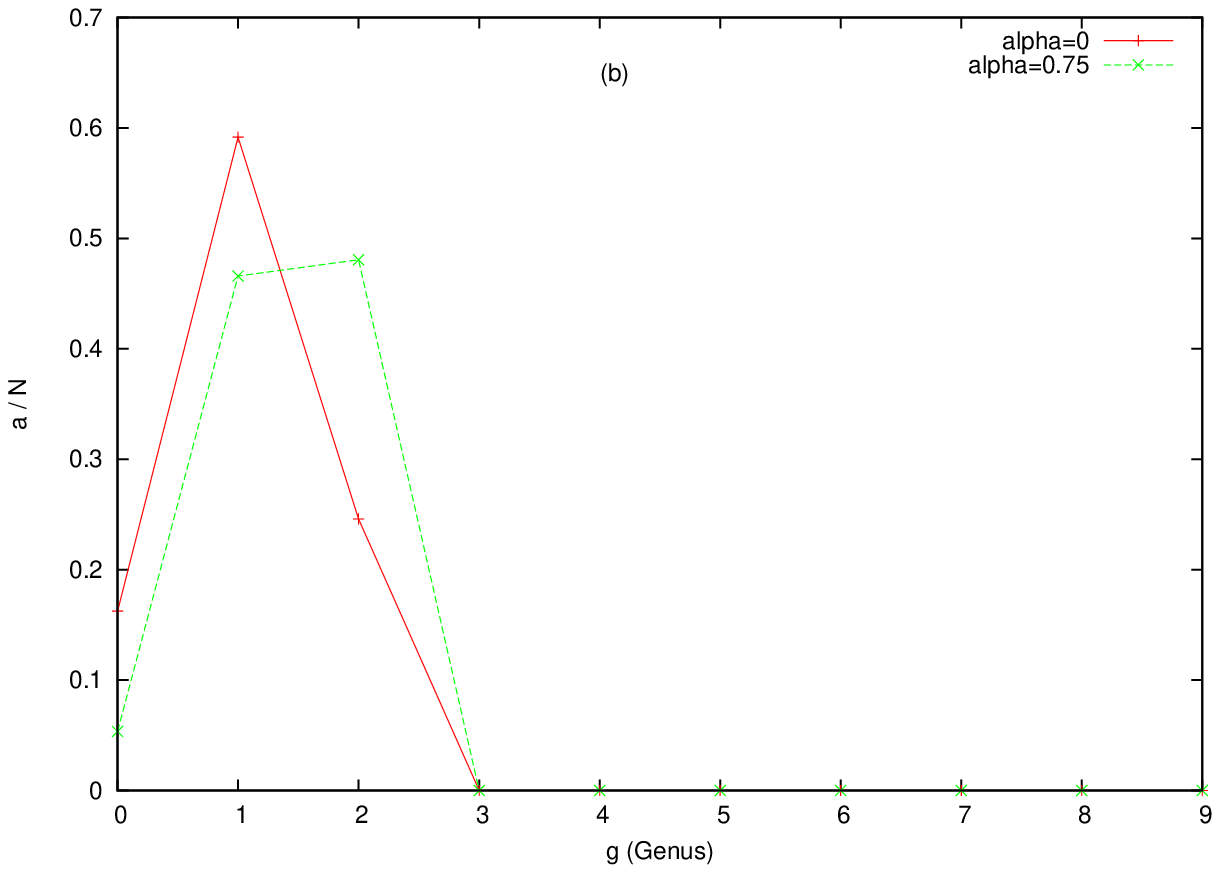}
\includegraphics*[width=7cm]{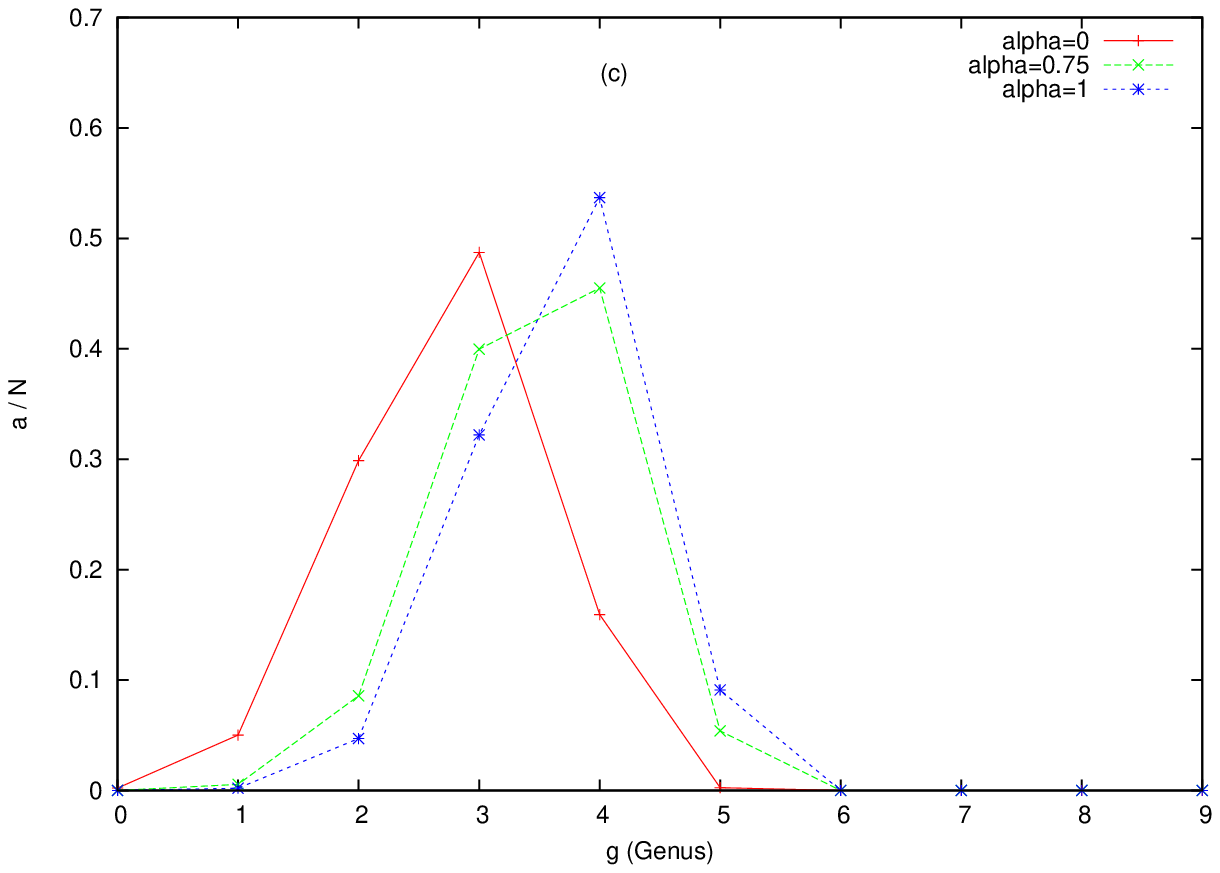}
\includegraphics*[width=7cm]{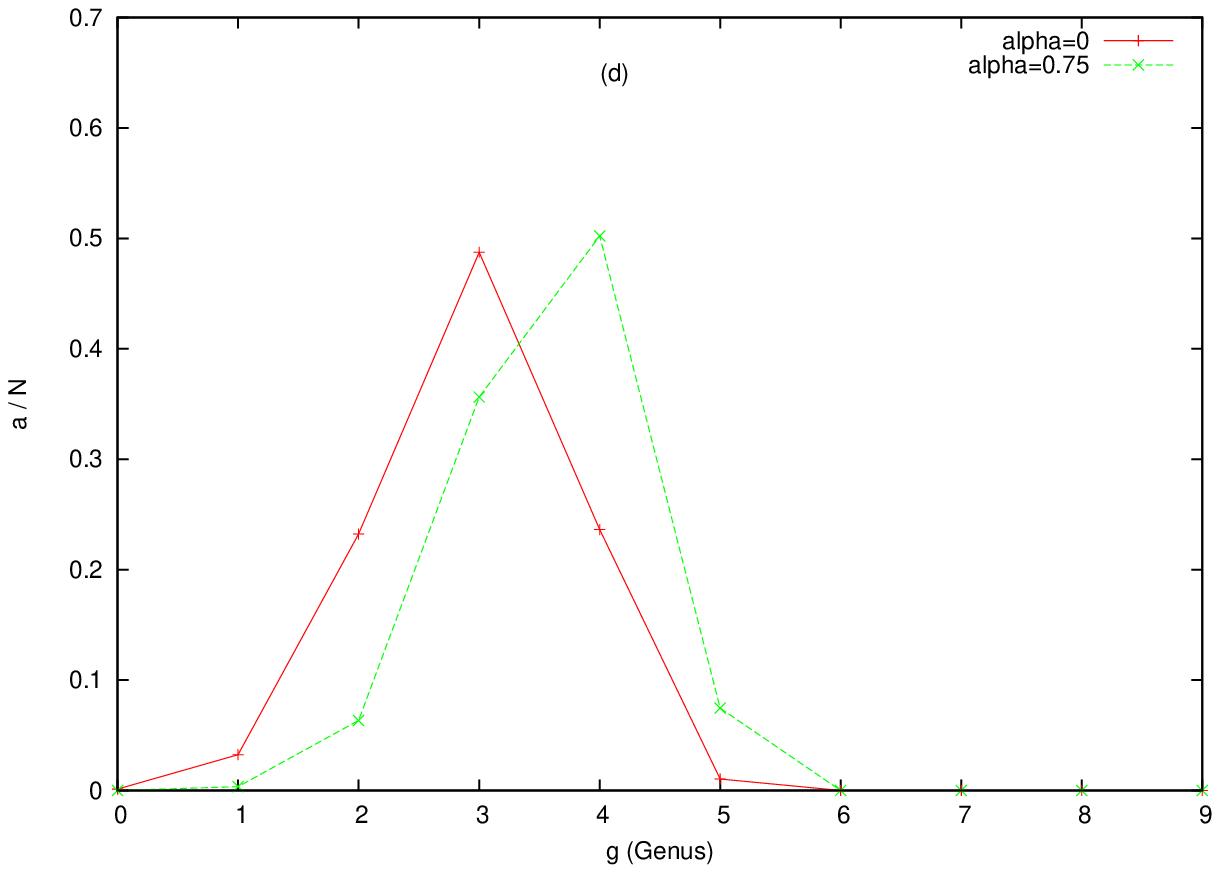}
\includegraphics*[width=7cm]{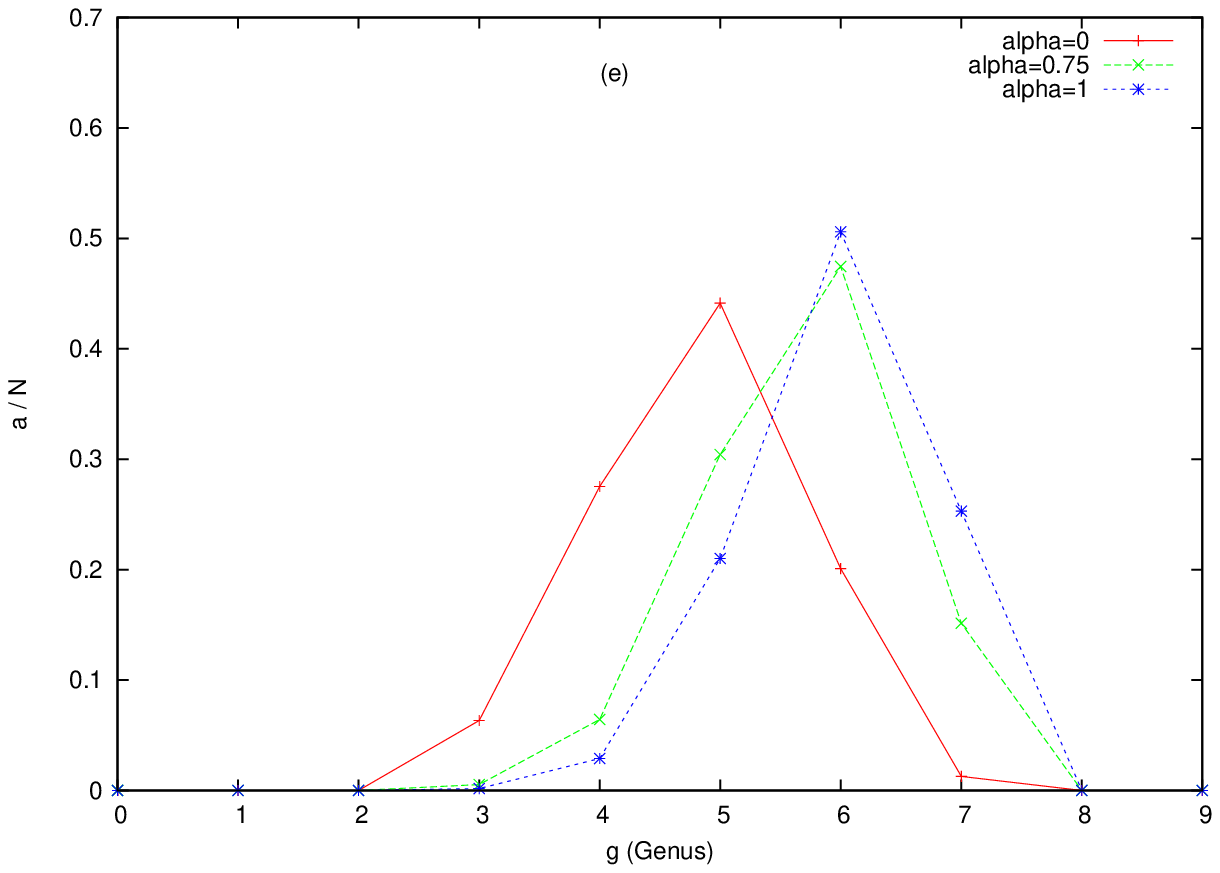}
\includegraphics*[width=7cm]{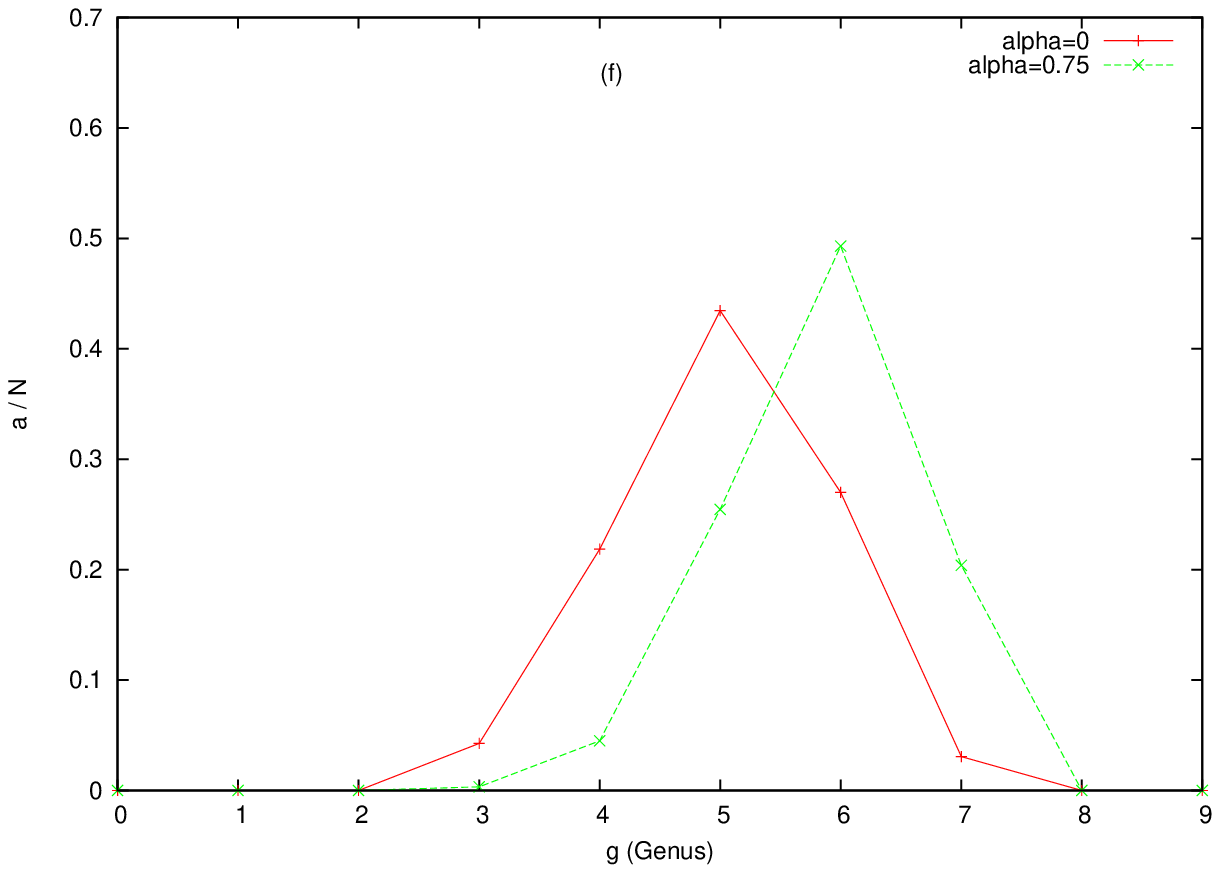}
\caption{Genus Distributions for (a). L=10, (b). L=11, (c). L=20,
(d). L=21, (e). L=30 and (f). L=31 with $\alpha(=0, 0.75, 1)$ values
are shown in these figures. For odd lengths ($L = 11, 21, 31$), the
$\alpha = 1$ distribution is absent. $\frac{a}{N}$ in the figures
should be understood as $\frac{a_{L, g, \alpha}}{{\cal
N}_{\alpha}}$.}
\end{figure*}

We compare the genus distributions for the following three pairs of
successive even and odd lengths: (i). $L = (10, 11)$ (Figures 3(a),
3(b) respectively), (ii). $L = (20, 21)$ (Figures 3(c), 3(d)
respectively) and (iii). $L = (30, 31)$ (Figures 3(e), 3(f)
respectively), for different $\alpha$ values (Figure 3). For a
chosen length, plots for different $\alpha=0,0.75,1$ are compared.
In the figure with odd length compared to the figure with even
length, it is observed that the $\alpha = 1$ curve is absent. This
is due to the absence of the partition function for odd lengths at
$\alpha=1$. It is also observed that the curve corresponding to
$\alpha = 0.75$ at $L = 11$, comprises of points which are an
average of the points on the $\alpha=0.75$ and $\alpha=1$ curves for
$L = 10$ (same genus points for the two lengths are considered). For
example in Figure 3(a) (for $L = 10$), the points corresponding to
$g = 2$ for $\alpha = 0.75$ and $\alpha = 1$ curves are averaged and
the $a_{L, g, \alpha}/{\cal N}_{\alpha}$ value thus obtained is
similar to the $a_{L, g, \alpha}/{\cal N}_{\alpha}$ value for $g =
2$ on the $\alpha = 0.75$ curve in Figure 3(b) (for $L = 11$). The
same is observed on comparison of other lengths ($L = (20, 21)$ and
$L = (30, 31)$). The shape of the $\alpha = 0.75$ curve for the odd
length figure (L=11) seems to take the shape of $\alpha = 1$ curve
in the even length figure(L=10). This is also observed in the figure
$a_{L, g, \alpha}/{\cal N}_{\alpha}$ Vs L for different genii
(g=0,1,2,3) when $\alpha=0,0.75,1$ are plotted together (Figure 4).
Notice that in the curve for $\alpha=0.75$ (Figure 4) for each
even-odd length, the points lie close together separated by large
distances from the nearest neighboring even-odd lengths.

This analysis extracts the otherwise not so visible differences in
the (i). total number of structures and (ii). the shape of genus
distributions for different $\alpha$'s as compared to the
distributions in RNA-MM \cite{1}. A study of the genus distributions
approaching $\alpha=1$ is particularly important as dramatic changes
take place as shown in Figures 3 and 4.

\begin{figure*}
\includegraphics*[width=7cm]{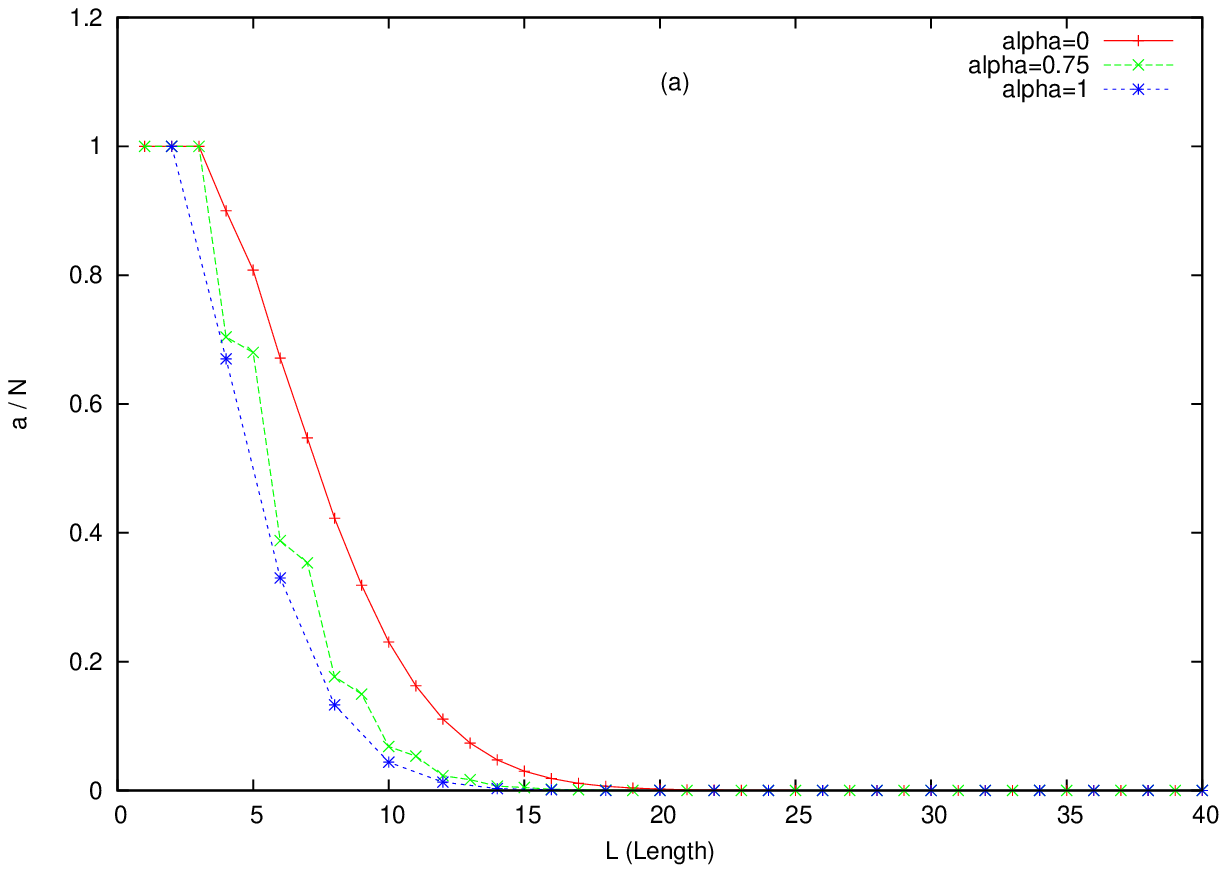}
\includegraphics*[width=7cm]{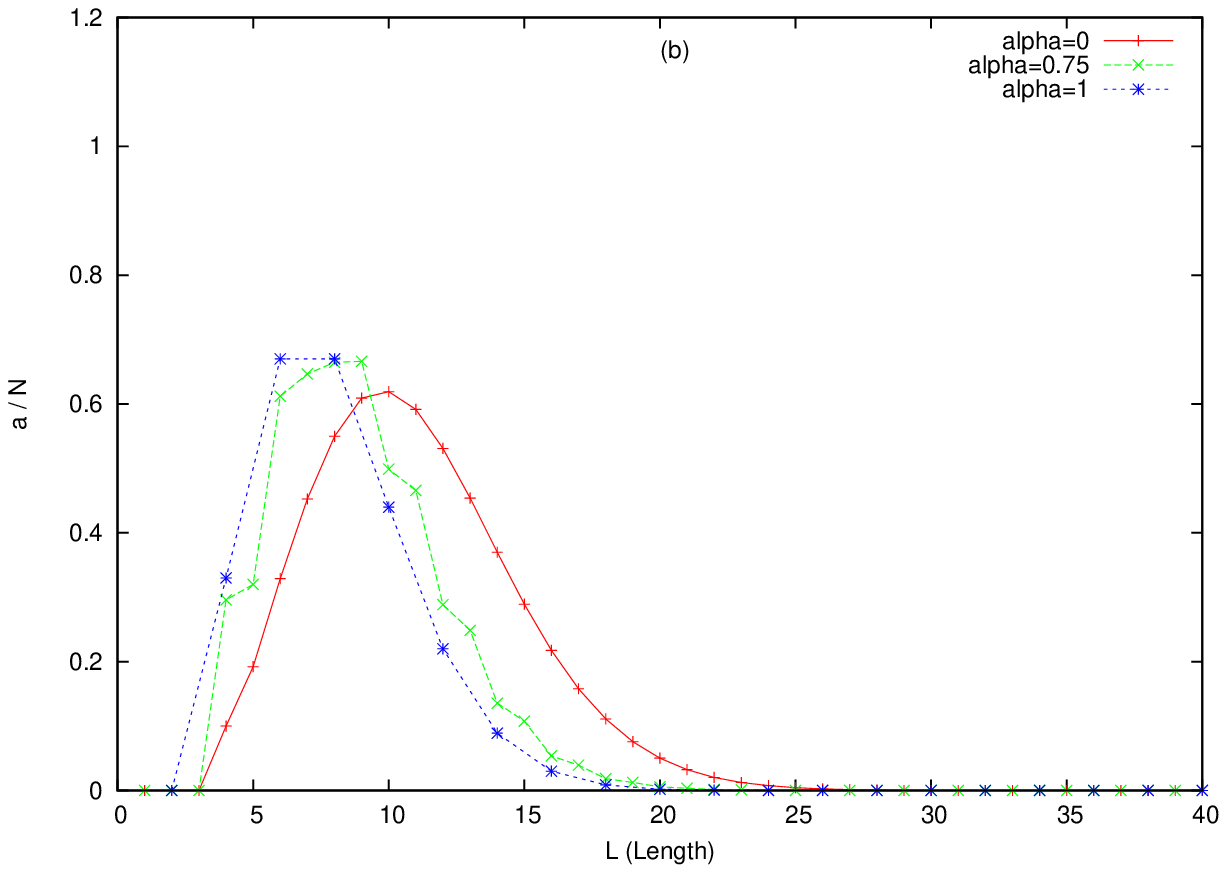}
\includegraphics*[width=7cm]{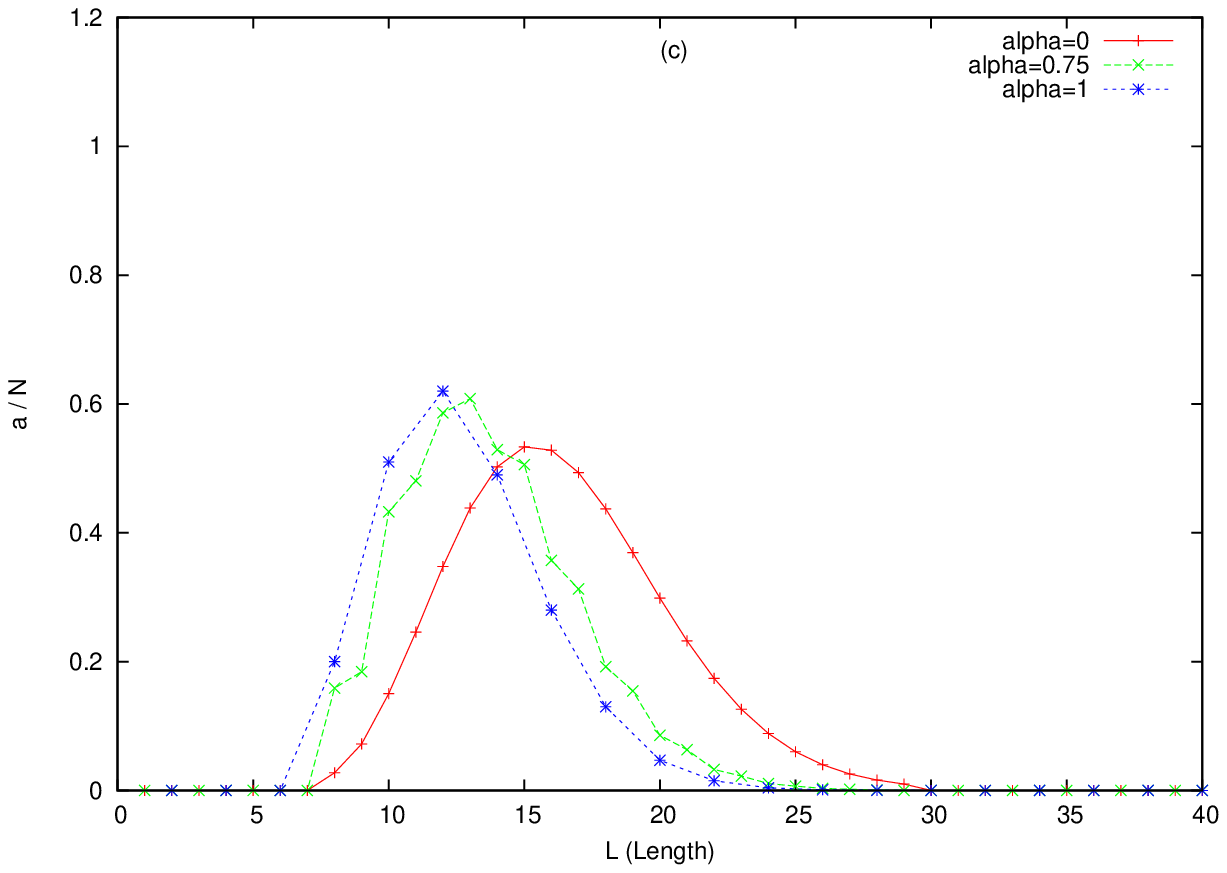}
\includegraphics*[width=7cm]{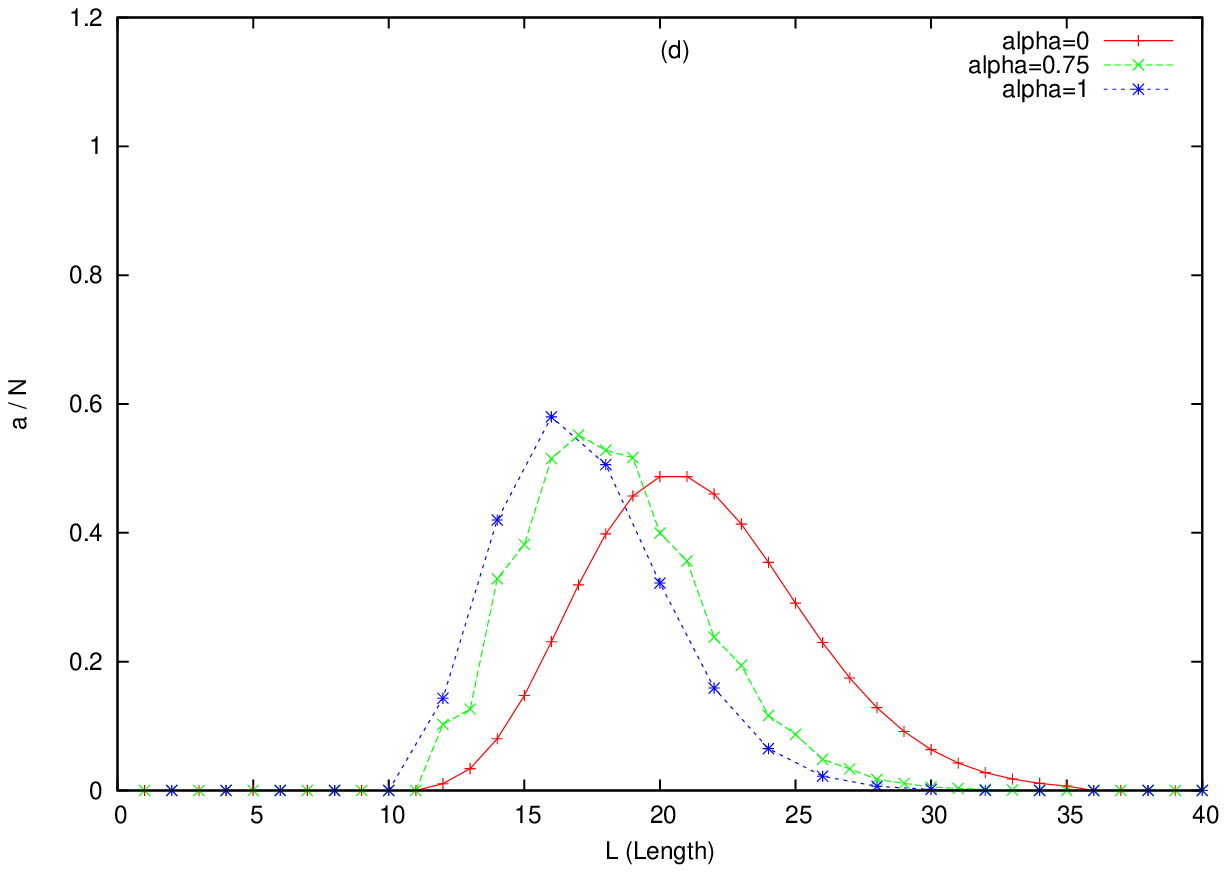}
\caption{Genus Distributions for (a). g=0, (b). g=1, (c). g=2 and
(d). g=3 with $\alpha(=0, 0.75, 1)$ values are shown in these
figures. $\frac{a}{N}$ in the figures should be understood as
$\frac{a_{L, g, \alpha}}{{\cal N}_{\alpha}}$.}
\end{figure*}

\section{Conclusions}
\label{5.}

We have given a detailed analytic calculation of the partition
  function after introducing a linear perturbation in the action
  of the nucleotide chain for the matrix model of RNA folding \cite{1} (section 2).
  A parameter $\alpha(=v/w)$ is found explicitly in the extended matrix model (RNA-EMM).
  The parameter $\alpha$ can be tuned to different
  values to give several combinations of the strength ratio of the
  original and the perturbed term thus giving various extensions of the model.
  For instance, for $\alpha=0$, the extended matrix model (RNA-EMM) reduces to the RNA-MM in \cite{1}. We find for
  $\alpha=1$, the partition function for odd lengths of
  the polymer chain vanishes completely. This brings significant
  changes in the genus distributions in terms of the number of structures and
  the shape of the distribution as compared with the RNA-MM in \cite{1}.
  We have compared the genus distributions at different lengths and genus
  for $\alpha = 0,0.75, 1$. The curve corresponding to $\alpha = 0.75$ at
  odd lengths comprises of points which are an average of the points on the
  $\alpha=0.75$ and $\alpha=1$ curves on the preceding plot of even length (the same genus
  for the two lengths is considered for comparison). Also, the shape of
  the $\alpha = 0.75$ curve for the odd lengths seem to
  take the shape of $\alpha = 1$ curve in the plot of even length (section 4).
  For other values of $\alpha$ we have different results for the partition
  function (listed in Table 1). The general form of $Z_{L, \alpha}(N)$ for the extended matrix model
  consists of an additional multiplicative factor of $(1-\alpha)$ with each term as compared
  to \cite{1}. The maximum value of genus g for the extended matrix model for even
  L and all $\alpha$ is found to be $L/4$ i.e. $g_{max}$ $\leq$ \hspace{0.2cm}$L/4$
  (found numerically). For odd lengths and $\alpha \neq 1$ this
  result is also true and for $\alpha=1$ there are no diagrams.

The diagrammatic representation of the extended matrix model is
  discussed in section 3. This follows from the general form of $Z_{L,
  \alpha}(N)$ for the extended matrix model found in section 2. The
  Feynmann diagrams corresponding to length $L=4$ of the
  polymer chain are shown in Figure 1. Each unpaired base in the polymer chain is
  associated with a factor of $(1-\alpha)$. Hence for odd lengths of the polymer chain when
  $\alpha = 1$ the partition function vanishes. $Z_{L, \alpha=1}(N)=0$ for odd lengths implies that no unpaired
  base is allowed for the $\alpha=1$ phase. For other values of $\alpha$, the unpaired bases in the
  chain are weighted by a factor of $(1-\alpha)$. We define a ``structural
  transition'' where structures change from a ``unpaired-paired base phase'' to a ``completely paired base phase''
  at $\alpha = 1$ (section 4). The perturbation has thus created a phase where RNA
  structures with a limited biological activity (as only structures
  with paired bases are possible) are separated out from the otherwise
  possible vast number of structures (where structures with both paired and unpaired bases are possible).

Our future work will concentrate on a study of more general and
  realistic perturbations and an analysis of their genus distributions.
  Another interesting aspect of the problem is to take into
  consideration Watson-Crick and Wobble (G-U) pairings in the interaction ($V_{i, j}$) between
  the nucleotides so that the condition of a base pairing
  with any other base (in models discussed so far) is relaxed.
  This will be a more realistic approach towards the problem of RNA
  folding but it is not a trivial problem. We would like to study
  these generalized matrix models with more mathematical rigour as the addition
  of a linear term in the action can produce non-trivial results as shown
  in \cite{11,12} for similar matrix models. The analysis of asymptotics for
  the extended matrix model (RNA-EMM) will also be dealt with in a future work.

\noindent{\bf Acknowledgements}

In section 3 which discusses the diagrammatic representation for the
extended matrix model is inspired by the comments of the referee of
Nuclear Physics B. We would like to thank Prof. H. Orland and G.
Vernizzi for e-discussions. We would like to thank CSIR Project No.
$03(1019)/05/EMR-II$ for financial support.

\end{document}